\documentclass{article}
\usepackage[utf8]{inputenc}
\usepackage[T1]{fontenc}
\usepackage{authblk}
\usepackage{dsfont}
\usepackage{geometry}
\usepackage{amsmath}
\usepackage{amssymb}
\newgeometry{tmargin=3cm, bmargin=3cm, lmargin=2cm, rmargin=2cm}
\title{Kantian equilibria in classical and quantum symmetric games}
\author{Piotr Frąckiewicz}
\affil{\small{Institute of Exact and Technical Sciences, \\ Pomeranian University in S\l upsk, Poland;\\ piotr.frackiewicz@apsl.edu.pl}}
\date{April 2021}
\newtheorem{definition}{Definition}
\newtheorem{theorem}{Theorem}
\newtheorem{example}{Example}
\newtheorem{lemma}{Lemma}
\newtheorem{proposition}{Proposition}
\newtheorem{corollary}{Corollary}
\newenvironment{proof}{\noindent\textit{Proof~~}}
{\nolinebreak[4]\hfill$\blacksquare$\\\par}
\DeclareMathOperator*{\argmax}{\arg\!\max}
\begin{document}
\maketitle
\begin{abstract}
    The aim of the paper is to examine the notion of simple Kantian equilibrium in $2 \times 2$ symmetric games and their quantum counterparts. We focus on finding the Kantian equilibrium strategies in the general form of the games. As a result, we derive a formula that determines the reasonable strategies for any payoffs in the bimatrix game. This allowed us to compare the payoff results for classical and quantum way of playing the game. We showed that a very large part of $2\times 2$ symmetric games have more beneficial Kantian equilibria when they are played with the use of quantum strategies.
\end{abstract}
\section{Introduction}
Game theory, launched in 1928 by John von Neumann in a paper \cite{neumann} and developed in 1944 by John von Neumann and Oskar Morgenstern in a book \cite{neumann2} is one of the branches of applied mathematics. The aim of this theory is mathematical modeling of behavior of rational players in conflict situations. The players are assumed to strive to maximizing their own benefit and take into account all possible ways of behaving of the other players.

Within this theory new ideas are still proposed. One of the latest trends is to define and study new solution concepts. A fundamental concept used for predicting players' behavior is Nash equilibrium \cite{nash}. It defines a strategy vector at which no player has a profitable deviation from a strategy of that vector. Beside of that kind of stability Nash equilibrium always exists in finite games. Therefore, the use of Nash equilibria is a first step in finding reasonable moves of the players. Nash equilibria may often indicate non-optimal solutions. Moreover, a game may have multiple Nash equilibria that imply different outcomes.  Then one can use refinements of Nash equilibria that impose additional restrictions on strategy vectors \cite{myerson}.

In the case of many games like the Prisoner's Dilemma game, Nash equilibria may imply very low payoffs compared to other payoffs available in the game. This undoubtedly have had an impact on the promotion of non-Nash equilibrium based solution concepts. There is a significant number of articles devoted to solution concepts that do not derive from Nash equilibra. One of the best references here is \cite{fourny}. The paper introduces the Perfectly Transparent Equilibrium - the idea based on rounds elimination of strategy profiles that do not imply maximin payoffs. Another notion worth mentioning is Berge equilibrium \cite{berge, colman, pbf}. The concept is based on altruism in the sense that each player's aim is to maximize the payoffs of the other players.

The subject of our work is the notion of Kantian equilibrium \cite{roemerbook}. It follows from a line of reasoning suggested by Hofstadter \cite{hofstadter}. His idea assumes that the players are superrational. This means that they are rational and also they are able to conduct a meta-analysis taking into account that the other players has the same reasoning. In the case of a symmetric game, the players have exactly the same strategic position. So, if one player predicts a rational strategy, the other players should also come up with the identical strategy. Examining profiles consisting of the same strategies makes finding Kantian equilibria a lot easier compared to Nash equilibria. The task is actually to maximize a payoff function with respect to a strategy of one of the players. This is particularly relevant in studying a quantum game in which a player's unitary strategy depends even on three parameters. 

Quantum game theory is a field developed on the border of game theory and quantum information. This is an interdisciplinary area of research. It assumes that games are played with the aid of objects that behave according to the laws of quantum mechanics. The theory was initiated with considering a simple extensive-form game in \cite{meyer}. D. Meyer showed that a player equipped with unitary operators has a winning strategy. Another fundamental paper on quantum games is \cite{eisert}. The scheme defined by J. Eisert, M. Wilkens and M. Lewenstein was the first formal protocol of playing a general $2\times 2$ game. According to \cite{eisert}, players' strategies are unitary operators that depend on two parameters. These operators are performed on a maximally entangled two-qubit state. L. Marinatto and T. Weber introduced an alternative model of playing a quantum game. In their scheme for a $2\times 2$ game players' strategies are restricted to two unitary operators (the identity and the Pauli operator $X$). The operators are performed on a fixed two-qubit state (not necessarily entangled). Quantum game theory also includes quantum models with infinite strategy sets. A minimalistic model of quantum Cournot duopoly introduced by H. Li, J. Du and S. Massar in \cite{lidumassar} and generalized in \cite{fraduopol}. 


Our work focuses on Kantian equilibria in $2\times 2$ symmetric game and its quantum counterpart. We generalize the previous findings presented in \cite{roemerbook} by deriving the general formula for Kantian equilibria. We then examine this solution concept with respect to the Eisert-Wilkens-Lewenstein quantum approach to the game. 
\section{Preliminaries}
In this section, we review relevant notions from classical and quantum game theory that are needed to follow our work.

The basic model of games studied in game theory is a game in strategic form.
\begin{definition}\textup{\cite{maszler}}~
A game in strategic form (or in normal form) is an ordered triple $(N, (S_{i})_{i\in N}, (u_{i})_{i\in N})$, in which
\begin{itemize}
\item $N = \{1,2, \dots, r\}$ is a finite set of players.
\item $S_{i}$ is the set of strategies of player $i$, for every player $i\in N$.
\item $u_{i}\colon S_{1}\times S_{2} \times \cdots \times S_{r} \to \mathds{R}$ is a function associating each vector of strategies $s= (s_{i})_{i\in N}$ with the payoff $u_{i}(s)$ to player $i$, for every player $i\in N$.
\end{itemize}
\end{definition}
A game in strategic form proceeds in the following way. Each player $i \in N$ chooses one of her strategies $s_{i}\in S_{i}$. In this way, the players determine a strategy vector $(s_{1}, s_{2}, \dots, s_{r})$. Then, for each player $i$, the payoff function $u_{i}$ determines a payoff $u_{i}(s_{1}, s_{2}, \dots, s_{r})$. 

A player can also choose their own strategies according to a probability distribution. Then we say that she plays a mixed strategy. Formally, a mixed strategy is an element of the set \cite{maszler}
\begin{equation}
    \Sigma_{i} = \left\{\sigma_{i}\colon S_{i} \to [0,1]\colon \sum_{s_{i} \in S_{i}}\sigma_{i}(s_{i}) = 1\right\}.
\end{equation}

In particular case, if a strategic-form game has two players and each player has two strategies, the game can be written as a $2\times 2$ matrix in which each element is a pair of real numbers
\begin{equation}\label{2x2g}
 (A,B) = \begin{pmatrix}
     (a_{00}, b_{00}) & (a_{01}, b_{01})\\
     (a_{10}, b_{10}) & (a_{11}, b_{11})
    \end{pmatrix}.
\end{equation}
Player 1's strategies are identified with the rows and player 2's strategies are identified with the columns. Players' mixed strategies in the case of game (\ref{2x2g}) will be denoted by $(p,1-p)$ and $(q,1-q)$, respectively. Then the expected payoff resulting from playing the mixed strategies is 
\begin{equation}
    \begin{aligned}
    &u_{1}((p,1-p), (q,1-q)) = pqa_{00} + p(1-q)a_{01} + (1-p)qa_{10} + (1-p)(1-q)a_{11}, \\ 
    &u_{2}((p,1-p), (q,1-q)) = pqb_{00} + p(1-q)b_{01} + (1-p)qb_{10} + (1-p)(1-q)b_{11}.
    \end{aligned}
\end{equation}
Symmetry is common in two-player normal form games. It is particularly visible in $2\times 2$ bimatrix games discussed in many game theory textbooks. Games such as the Prisoner's Dilemma, Chicken or Stag Hunt are the examples of symmetric games. Informally, one can say that a symmetric game is one that looks the same for all the players \cite{binmore}. A more precise definition is as follows \cite{osborne, plan}:
\begin{definition}\textup{\cite{plan}}~
Let $N=\{1,2\}$ and $S_{1} = S_{2}$. A game $\Gamma = (N, (S_{1}, S_{2}), (u_{1}, u_{2}))$ is symmetric if for all pairs of strategies $(x,y) \in S_{1}\times S_{2}$
\begin{equation}\label{symmetric}
    u_{1}(x,y) = u_{2}(y,x).
\end{equation}
\end{definition}
\noindent In the case of a finite symmetric two-player game, condition (\ref{symmetric}) means that $B = A^{T}$ in (\ref{2x2g}). Then a symmetric $2\times 2$ bimatrix game takes the following form:
\begin{equation}
\begin{pmatrix}\label{sym2x2}
     (a_{00}, a_{00}) & (a_{01}, a_{10})\\
     (a_{10}, a_{01}) & (a_{11}, a_{11})
    \end{pmatrix}.
\end{equation}

The Eisert-Wilkens-Lewenstein (EWL) scheme \cite{eisert} has undoubtedly been one of the most used scheme for quantum games. In the EWL scheme, players' strategies are unitary operators that each of two players acts on a maximally entangled state. In the literature there are a few descriptions of the EWL scheme that are strategically equivalent. In what follows, we present a concise form that we adapted for the purpose of my research.

Let
\begin{equation}
|\Psi\rangle = \frac{|00\rangle + i|11\rangle}{\sqrt{2}}, \quad C_{0} = \begin{pmatrix}
 1 & 0 \\
 0 & 1
\end{pmatrix}, \quad C_{1} = \begin{pmatrix}
 0 & i \\
 i & 0
\end{pmatrix}.
\end{equation}
For $k,l \in \{0,1\}$ define
\begin{equation}
|\Psi_{kl}\rangle = C_{k}\otimes C_{l} |\Psi\rangle. 
\end{equation}
Then $\{|\Psi_{kl}\rangle \colon k,l \in \{0,1\}\}$ is a basis for $\mathds{C}^2\otimes \mathds{C}^2$.
\begin{definition}\label{ewldef}
The Eisert-Wilkens-Lewenstein approach to game (\ref{2x2g}) is defined by a triple $(N, (D_{i})_{i\in N}, (v_{i})_{i\in N})$, where 
\begin{itemize}
    \item $N = \{1,2\}$ is the set of players,
    \item $D_{i}$ is a set of unitary operators from $\mathsf{SU}(2)$ with typical element 
    \begin{equation}
        U_{i}(\theta_{i}, \alpha_{i}, \beta_{i}) = \begin{pmatrix}
      e^{i\alpha}\cos\frac{\theta}{2} & ie^{i\beta}\sin\frac{\theta}{2} \\ 
      ie^{-i\beta}\sin\frac{\theta}{2} & e^{-i\alpha}\cos\frac{\theta}{2}
      \end{pmatrix}, \quad \theta_{i} \in [0,\pi], \alpha, \beta \in [0,2\pi),
    \end{equation}
    \item $v_{i}\colon D_{1}\otimes D_{2} \to \mathds{R}$ is player $i$'s payoff function given by
    \begin{align}
        &v_{1}(U_{1}\otimes U_{2}) = \sum^1_{k,l=0}a_{kl}|\langle \Psi_{kl}|U_{1}\otimes U_{2}|\Psi\rangle|^2\\
        &v_{2}(U_{1}\otimes U_{2}) = \sum^1_{k,l=0}b_{kl}|\langle \Psi_{kl}|U_{1}\otimes U_{2}|\Psi\rangle|^2,
    \end{align}
    where $a_{kl}$ and $b_{kl}$ for $k,l \in \{0,1\}$ are the payoffs of (\ref{2x2g}).
\end{itemize}
\end{definition}
According to Definition~\ref{ewldef}, the EWL approach to (\ref{2x2g}) can be regarded as a normal-form game in which the strategies are unitary operators that the players perform on the state $|\Psi\rangle$. The payoff function is then the expected value of measurement on the final state, 
\begin{equation}
U_{1}\otimes U_{2}|\Psi\rangle = \sum^1_{k,l = 0}\langle \Psi_{kl}|U_{1}\otimes U_{2}|\Psi\rangle|\Psi_{kl}\rangle
\end{equation}
with respect to the basis $\{|\Psi_{kl}\rangle \colon k,l \in \{0,1\}\}$.

One of the main features of the von Neumann-Morgenstern utility says that if player $i$'s preferences in a game are represented by the expected value of payoffs, then every positive affine transformation of the payoffs also represents these preferences.
\begin{definition}\textup{\cite{maszler}}
Let $u\colon X \to \mathds{R}$ be a function. A function $v\colon X \to \mathds{R}$ is a positive affine transformation of $u$ if there exists a positive real number $\alpha >0$ and a real number $\beta$ such that for each $x\in X$
\begin{equation}
    v(x) = \alpha u(x) + \beta.
\end{equation}
\end{definition}
\begin{theorem}\textup{\cite{maszler}}\label{th1}
If $u_{i}$ is a linear utility function representing player $i$'s preferences, then every positive affine transformation of $u_{i}$ is also a linear utility function representing the preferences. 
\end{theorem}
The next example illustrates Theorem~\ref{th1}.
\begin{example}
Let us consider the following bimatrix game:
\begin{equation}\label{nonzerogame}
    \begin{pmatrix}
    (-14, 15) & (-2,-3) \\ 
    (-4,0) & (12,12)
    \end{pmatrix}.
\end{equation}
If we transform player 1 and 2's payoffs by the positive affine transformations $\frac{1}{2}x + 5$ and $\frac{1}{3}x - 3$, respectively, we obtain
\begin{equation}\label{zerogame}
    \begin{pmatrix}
 (-2,2) & (4,-4) \\
 (3,-3) & (-1,1)
    \end{pmatrix}.
\end{equation}
Although (\ref{zerogame}) is a zero-sum game in contrast to (\ref{nonzerogame}), both games are equivalent with respect to players' preferences about the result of the game. Both games have the unique Nash equilibrium $((2/5, 3/5), (1/2, 1/2))$.  
\end{example}
More generally, if $(p_{i})^m_{i=1}$ is a probability distribution over player $i$'s payoffs $(a_{i})^m_{i=1}$ and $\sum^m_{i=1} p_{i}a_{i}$ is the expected payoff then for $\alpha \in \mathds{R}_{+}$ and $\beta \in \mathds{R}$
\begin{equation}
    \alpha \sum^m_{i=1} p_{i}a_{i} +\beta = \sum^m_{i=1}\alpha p_{i}a_{i} + \sum^m_{i=1}\beta p_{i} = \sum^m_{i=1}(\alpha a_{i} + \beta)p_{i}.
\end{equation}
Hence, if a player prefers a probability distribution $(p_{i})^m_{i=1}$ over $(p'_{i})^m_{i=1}$ then 
\begin{equation}
    \sum^m_{i=1} p_{i}a_{i} \geq \sum^m_{i=1} p'_{i}a_{i} \Leftrightarrow \sum^m_{i=1}(\alpha a_{i} + \beta)p_{i} \geq \sum^m_{i=1}(\alpha a_{i} + \beta)p'_{i}.
\end{equation}
Since the expected payoff in the EWL scheme is also in the form of $\sum^m_{i=1} p_{i}a_{i}$, where $a_{i}$ are the payoffs in the classical game, a positive affine transformation of the classical game does not change players' preferences in the associated EWL game. 
\section{Kantian equilibria in $2\times2$ symmetric games}
The definition of Kantian equilibrium varies according to a type of game \cite{roemer,  istrate}. In what follows, we reproduce the one concerning a mixed extension of a finite normal-form game. 

Let us consider a normal-form game $(N, (S_{i})_{i\in N}, (u_{i})_{i\in N})$ that have identical strategy sets, i.e., $S_{1} = S_{2} = \dots = S_{n} = S$ and let $\Delta(S)$ be the set of probability distributions on $S$.
\begin{definition}\label{SKE}
A simple Kantian equilibrium (SKE) is a vector $(\tau^*, \tau^*, \dots, \tau^*) \in \Delta(S)^n$ such that 
\begin{equation}
    \tau^* \in \argmax_{\tau \in \Delta(S)}{u_{i}(\tau, \tau, \dots, \tau)}.
\end{equation}
\end{definition}
Obviously, Definition~\ref{SKE} can be easily modified when one considers pure strategies $S$ or quantum strategies (i.e., unitary operators) instead of $\Delta(S)$.

In \cite{roemerbook} simple Kantian equilibria are found for a few examples of $2\times2$ symmetric games. 
Our results generalize that of \cite{roemerbook}. We provide a concise formula for a general $2\times 2$ symmetric game. 

Let us consider a two-player symmetric game (\ref{sym2x2}). 
Let us first consider the case $a_{00} \ne a_{11}$. 
There is no loss of generality in assuming that $a_{00} > a_{11}$. 
To simplify Kanitan equilibrium analysis let us apply a positive affine transformation to (\ref{sym2x2}) in the following form:
\begin{equation}
    f(x) = \frac{1}{a_{00} - a_{11}}(x-a_{11}).
\end{equation}
Then, game (\ref{sym2x2}) is transformed into a preference-equivalent game
    \begin{equation}\label{sym10}
\begin{pmatrix}
     (1, 1) & (a, d-a)\\
     (d-a, a) & (0, 0)
    \end{pmatrix}, 
\end{equation}
where 
\begin{equation}\label{sub1}
    a = \frac{a_{01}-a_{11}}{a_{00}-a_{11}}, ~ d = \frac{a_{01}+ a_{10} - 2a_{11}}{a_{00} - a_{11}}.
\end{equation}
Let us first determine the expected payoff $u_{i}((p, 1-p),(p,1-p))$ of player $i$ resulting from playing a strategy vector $((p, 1-p),(p,1-p))$ in (\ref{sym10}). We obtain
\begin{equation}\label{equ1u2}
    u_{1}((p, 1-p),(p,1-p)) = u_{2}((p, 1-p),(p,1-p)) = p^2 + d(1-p)p, \quad p\in [0,1].
\end{equation}
By Definition~\ref{SKE}, simple Kantian equilibria in (\ref{sym10}) are determined by points that maximize (\ref{equ1u2}). We shall consider two cases. If $d>1$, then the point $p^* = -d/(2(1-d))$ is a local maximum point of $x^2+d(1-x)x, ~x\in \mathds{R}$. It maximizes (\ref{equ1u2}) if $0\leq p^* \leq 1$ or $d\geq 2$. For $1<d<2$,
\begin{equation}
    \frac{d}{dp}u_{i}((p, 1-p),(p,1-p)) = 2p(1-d) + d > 0.
\end{equation}
Hence, $p^* =1$ maximizes (\ref{equ1u2}) for $1<d<2$. 

If $d \leq 1$, function (\ref{equ1u2}) attains its maximum at one of the endpoints of $[0,1]$. In this case, it is the point $p^*=1$. 
Summarizing, we have thus proved the following lemma:
\begin{lemma}\label{lemma1}
In a symmetric $2\times 2$ game in the form of (\ref{sym10}), if $u_{i}((p,1-p),(q,1-q))$ is player $i$'s payoff function, then it follows that 
\begin{equation}\label{formula1}
\argmax_{p\in [0,1]} u_{i}((p,1-p),(p,1-p))= \begin{cases}
\{1\} &\text{if}~ d<2,\\
\left\{\frac{-d}{2(1-d)}\right\} &\text{if}~d\geq 2.
\end{cases}
\end{equation}
\end{lemma}
Let us now consider bimatrix (\ref{sym2x2}) in which $a_{00} = a_{11}$. By using a positive affine transformation $g(x) = x-a_{00}$ we are left with the task of determining simple Kantian equilibria in 
\begin{equation}\label{sym00}
\begin{pmatrix}
     (0, 0) & (b, e-b)\\
     (e-b, b) & (0, 0)
    \end{pmatrix}, 
\end{equation}
where
\begin{equation}\label{sub2}
     b = a_{01}-a_{00}, ~ e = a_{01}+ a_{10} - 2a_{00}.
\end{equation}
Now, the problem of finding SKE comes down to determining the points that maximize 
\begin{equation}
    u_{i}((p,1-p),(p,1-p)) = ep(1-p),\quad p\in [0,1].
\end{equation}
We leave it to the reader to verify the following lemma:
\begin{lemma}\label{lemma2}
In a symmetric $2\times 2$ game in the form of (\ref{sym00}), if $u_{i}((p,1-p),(q, 1-q))$ is player $i$'s payoff function, then it follows that 
\begin{equation}\label{formula2}
    \argmax_{p\in [0,1]}u_{i}((p,1-p),(p,1-p)) = \begin{cases}
    \{0,1\} &\text{if}~e<0, \\
    [0,1] &\text{if}~e=0, \\
    \left\{\frac{1}{2}\right\} &\text{if}~e>0.
    \end{cases}
\end{equation}
\end{lemma}
Although formulae~(\ref{formula1}) and (\ref{formula2}) find simple Kantian equilibria for games~(\ref{sym10}) and (\ref{sym00}), respectively, Lemmas~\ref{lemma1} and \ref{lemma2} enable us to generalize the results to arbitrary symmetric $2\times 2$ games (\ref{sym2x2}).
\begin{proposition}\label{prop1}
Let $u_{i}((p,1-p),(q, 1-q))$ be a player $i$'s payoff function in symmetric $2\times 2$ game (\ref{sym2x2}) in which $a_{00} \geq a_{11}$. Then
\begin{equation}\label{propeq1}
    \argmax_{p\in [0,1]}u_{i}((p,1-p),(p,1-p)) = \begin{cases}
    \{1\} &\text{if}~~a_{01}+a_{10} - 2a_{00} \leq 0 ~~\text{and}~~ a_{00}>a_{11}, \\ 
    \{0,1\} &\text{if}~~a_{01}+a_{10} - 2a_{00} < 0 ~~\text{and}~~ a_{00}=a_{11},\\
    [0,1] &\text{if}~~a_{01}+a_{10} - 2a_{00} = 0 ~~\text{and}~~ a_{00}=a_{11}, \\ 
    \left\{\frac{a_{01} + a_{10} - 2a_{11}}{2(a_{01}+a_{10} - a_{00}-a_{11})}\right\} &\text{if}~~a_{01}+a_{10} - 2a_{00} > 0 ~~\text{and}~~ a_{00}\geq a_{11}.
    \end{cases}
\end{equation}
\end{proposition}
\begin{proof}
The first part of (\ref{propeq1}) follows from the first part of (\ref{formula1}), i.e., the condition $d<2$ is equivalent to $a_{01} + a_{10} - 2a_{00} < 0$ by (\ref{sub1}). Moreover, $-d/(2(1-d)) = 1$ for $d=2$.

Similarly, from the first and second part of (\ref{formula2}) we obtain the second and third part of (\ref{propeq1}).

Substituting the form of $d$ given in (\ref{sub1}) into $-d/(2(1-d))$ we obtain 
\begin{equation}\label{mix}
    p = \frac{a_{01} + a_{10} - 2a_{11}}{2(a_{01} +a_{10} - a_{00} - a_{11})}.
\end{equation}
In particular, $p$ given in (\ref{mix}) is equal to 1/2 for $a_{00} = a_{11}$.
\end{proof}
\begin{corollary}
Formula (\ref{propeq1}) in Proposition~\ref{prop1} also applies to game (\ref{sym2x2}) with $a_{00}<a_{11}$ by the reverse numbering of players' strategies, i.e., by swapping two rows and two columns in the bimatrix. Then we get a game that is isomorphic to (\ref{sym2x2}) and $a_{00}>a_{11}$. The value of $p$ resulting from (\ref{propeq1}) is then the probability of playing the second strategy in the initial game. 
\end{corollary}
SKE given by formula (\ref{propeq1}) implies the following payoff outcomes in a symmetric $2\times 2$ game: 
\begin{equation}\label{paygen}
    \max_{p\in [0,1]}u_{i}((p,1-p),(p,1-p)) = \begin{cases}
    a_{00} &\text{if}~~a_{01}+ a_{10} - 2a_{00} \leq 0, \\ 
    \frac{(a_{01}+a_{10})^2 - 4a_{00}a_{11}}{4(a_{01} + a_{10} - a_{00}-a_{11})} &\text{if}~~a_{01}+ a_{10} - 2a_{00} > 0.
    \end{cases}
\end{equation}
In what follows, we apply (\ref{propeq1}) and (\ref{paygen}) to concrete examples of bimatrix $2\times 2$ games.
\begin{example}\label{ex2}
\begin{enumerate}
    \item[1.] The general structure of the Prisoner's Dilemma game can be expressed by (\ref{sym2x2}) that satisfies 
    \begin{equation}\label{pd}
    a_{10}>a_{00}>a_{11}>a_{01} ~~\text{and}~~ a_{00}>\frac{a_{01}+a_{01}}{2}.
\end{equation}
The second condition of (\ref{pd}) is aimed at preventing the players from alternating between their first and second strategies. The condition coincides with the one of (\ref{propeq1}). This means that the simple Kantian equilibrium is $((p,1-p),(p,1-p)) = ((1,0),(1,0))$ with the payoff outcome $a_{00}$ for each player.
\item[2.] A symmetric game isomorphic to the Battle of the Sexes game can be described by (\ref{sym2x2}) such that
\begin{equation}
    a_{01} > a_{10} > a_{00} = a_{11}. 
\end{equation}
By (\ref{prop1}), the strategy of SKE is 
\begin{equation}
    p = \frac{a_{01}+a_{10}-2a_{00}}{2(a_{01}+a_{10}-2a_{00})} = \frac{1}{2}
\end{equation}
with the resulting payoff 
\begin{equation}
    \frac{(a_{01}+a_{10})^2-4a^2_{00}}{4(a_{01}+a_{10}-2a_{00})} = \frac{1}{4}(2a_{00} + a_{01} + a_{10}).
\end{equation}
\end{enumerate}
\end{example}
\section{Kantian equilibrum in the EWL-type quantum games}
One of the main motivation for studying quantum games is to search for reasonable quantum strategy profiles that would imply higher payoffs than ones implied by classical strategies. In this section we examine simple Kantian equilibria in the EWL quantum game to see whether players can benefit from playing quantum game.

First, we need to make sure that the game generated by the EWL scheme is symmetric so that we can apply the notion of simple Kantian equilibrium. 
\begin{lemma}
The Eisert-Wilkens-Lewenstein approach to a symmetric $2\times 2$ game is a symmetric game. 
\end{lemma}
\begin{proof}
Our proof starts with the observation that 
\begin{equation}\label{cztery}
    \begin{aligned}
    &\langle \Psi_{00}|U_{1}\otimes U_{2}|\Psi \rangle = \langle \Psi_{00}|U_{2}\otimes U_{1}|\Psi \rangle = \cos(\alpha_{1}+\alpha_{2})\cos\frac{\theta_{1}}{2}\cos\frac{\theta_{2}}{2} + \sin(\beta_{1}+\beta_{2})\sin\frac{\theta_{1}}{2}\sin\frac{\theta_{2}}{2}, \\
    &\langle \Psi_{01}|U_{1}\otimes U_{2}|\Psi \rangle = \langle \Psi_{10}|U_{2}\otimes U_{1}|\Psi \rangle = \cos(\alpha_{1}-\beta_{2})\cos\frac{\theta_{1}}{2}\sin\frac{\theta_{2}}{2} + \sin(\alpha_{2}-\beta_{1})\sin\frac{\theta_{1}}{2}\cos\frac{\theta_{2}}{2},\\
    &\langle \Psi_{10}|U_{1}\otimes U_{2}|\Psi \rangle = \langle \Psi_{01}|U_{2}\otimes U_{1}|\Psi \rangle = \cos(\alpha_{2}-\beta_{1})\sin\frac{\theta_{1}}{2}\cos\frac{\theta_{2}}{2} + \sin(\alpha_{1}-\beta_{2})\cos\frac{\theta_{1}}{2}\sin\frac{\theta_{2}}{2},\\
    &\langle \Psi_{11}|U_{1}\otimes U_{2}|\Psi \rangle = \langle \Psi_{11}|U_{2}\otimes U_{1}|\Psi \rangle = \cos(\beta_{1}+\beta_{2})\sin\frac{\theta_{1}}{2}\sin\frac{\theta_{2}}{2} - \sin(\alpha_{1}+\alpha_{2})\cos\frac{\theta_{1}}{2}\cos\frac{\theta_{2}}{2}.
\end{aligned}
\end{equation}
Since it is assumed that the $2\times 2$ game is symmetric then $a_{kl} = b_{lk}$ for $k,l \in \{0,1\}$. From (\ref{cztery}) it follows that
\begin{align}
    v_{1}(U_{1}\otimes U_{2}) &= \sum^1_{k,l=0}a_{kl}|\langle \Psi_{kl}|U_{1}\otimes U_{2}|\Psi\rangle|^2  = \sum^1_{k,l=0}a_{lk}|\langle \Psi_{kl}|U_{2}\otimes U_{1}|\Psi\rangle|^2 \nonumber \\&= \sum^1_{k,l=0}b_{kl}|\langle\Psi_{kl}|U_{2}\otimes U_{1}|\Psi\rangle|^2 = v_{2}(U_{2}\otimes U_{1}).
\end{align}
This finishes the proof.
\end{proof}
Let us first consider a game given by bimatrix (\ref{sym10}). The payoff functions in the EWL scheme associated with (\ref{sym10}) are symmetric. In particular, 
\begin{align}
    &|\langle \Psi_{00}|U^{\otimes 2}|\Psi\rangle|^2 = \left(\cos2\alpha \cos^2\frac{\theta}{2} + \sin2\beta \sin^2\frac{\theta}{2}\right)^2, \\ 
    &|\langle \Psi_{01}|U^{\otimes 2}|\Psi\rangle|^2 = |\langle \Psi_{10}|U^{\otimes 2}|\Psi\rangle|^2 = \left((\cos(\alpha - \beta)+ \sin(\alpha - \beta))\cos\frac{\theta}{2}\sin\frac{\theta}{2}\right)^2.
\end{align}
Therefore, for $i=1,2$, player $i$'s payoff function in the EWL game can be written as
\begin{equation}
    v_{i}(U,U) = |\langle \Psi_{00}|U^{\otimes 2}|\Psi\rangle|^2 + d|\langle \Psi_{01}|U^{\otimes 2}|\Psi\rangle|^2. 
\end{equation}
Obviously, the squared magnitudes of $|\langle \Psi_{ij}|U^{\otimes 2}|\Psi\rangle|^2$ sum to unity. Moreover, it is easy to check that 
\begin{equation}
    \max_{U \in \mathsf{SU}(2)}|\langle \Psi_{00}|U^{\otimes 2}|\Psi\rangle|^2 = 1, \quad \max_{U \in \mathsf{SU}(2)}|\langle \Psi_{01}|U^{\otimes 2}|\Psi\rangle|^2 = \frac{1}{2}.
\end{equation}
Note also that for $U' \in \argmax_{U\in \mathsf{SU}(2)}|\langle \Psi_{00}|U^{\otimes 2}|\Psi\rangle|^2$ we have $|\langle \Psi_{01}|U'^{\otimes 2}|\Psi\rangle|^2 = 0$. Similarly, for 
\begin{equation}
U'' \in \argmax_{U\in \mathsf{SU}(2)}|\langle \Psi_{01}|U^{\otimes 2}|\Psi\rangle|^2
\end{equation}
we see that $|\langle \Psi_{00}|U''^{\otimes 2}|\Psi\rangle|^2 = 0$. Hence, for $d>2$, 
\begin{equation}
    \max_{U\in \mathsf{SU}(2)}v_{i}(U,U) = \max_{U\in \mathsf{SU}(2)}d|\langle \Psi_{01}|U^{\otimes 2}|\Psi\rangle|^2 = \frac{d}{2}.
\end{equation}
Otherwise, $\max_{U\in \mathsf{SU}(2)}v_{i}(U,U) = 1$. 

The bimatrix (\ref{sym00}) can be handled in much the same way. The payoff function $v_{i}(U,U) = e|\langle \Psi_{01}|U^{\otimes 2}|\Psi\rangle|^2$ attains its maximum equal to $e/2$ if $e>0$, and $0$ otherwise. 

By using the inverse transformation $f^{-1}(y) = (a_{00}-a_{11})y + a_{11}$, $g^{-1}(y) = y+a_{11}$ and the substitutions (\ref{sub1}) and (\ref{sub2}), we can return to general payoffs to obtain $f^{-1}(d/2) = g^{-1}(e/2) = (a_{01}+a_{10})/2$ and $f^{-1}(1) = g^{-1}(0) = a_{00}$. Summarizing, we have thus proved the following proposition:
\begin{proposition}
Let $v_{i}(U,U)$ be a player $i$'s payoff function in the Eisert-Wilkens-Lewenstein quantum approach to symmetric $2\times 2$ game (\ref{sym2x2}) in which $a_{00} \geq a_{11}$. Then
\begin{equation}\label{propquantum}
    \max_{U\in \mathsf{SU}(2)}v_{i}(U,U) = \begin{cases}
    a_{00} &\text{if}~~a_{01}+a_{10} - 2a_{00} \leq 0, \\
    \frac{a_{01}+a_{10}}{2} &\text{if}~~a_{01}+a_{10} - 2a_{00} > 0.
    \end{cases}
\end{equation}
\end{proposition}
Comparing formulae (\ref{paygen}) and (\ref{propquantum}) for $a_{01}+a_{10} - 2a_{00} \leq 0$ we find that SKE implies the same payoff in both the classical and quantum game. For $a_{01}+a_{10} - 2a_{00} > 0$, SKE played in the quantum game results in a strictly better payoff than in the classical one. Indeed, 
\begin{equation}
\frac{a_{01} + a_{10}}{2} - \frac{(a_{01}+a_{10})^2 - 4a_{00}a_{11}}{4(a_{01}+a_{10} - a_{00}-a_{11})} = \frac{(a_{01}+a_{10} - 2a_{00})(a_{01}+a_{10}-2a_{11})}{4(a_{01}+a_{10}-a_{00}-a_{11})} >0.
\end{equation}
Let us reconsider the games from Example~\ref{ex2}. 
\begin{example}
\begin{enumerate}
    \item The Prisoner's Dilemma played with the use of the EWL scheme implies $a_{00}$--the result of SKE in the classical game. An example of a strategy of SKE is $U(\theta, \alpha, \beta) = U(0,\pi/2, 0)$.
    \item A symmetric game to isomorphic to the Battle of the Sexes game satisfies the condition $a_{01}+a_{10}-2a_{00} > 0$. Therefore, the payoff predicted by SKE is $(a_{01}+a_{10})/2$. Given particular payoffs in the Battle of the Sexes, e.g., $a_{01} = 5, a_{10} = 3, a_{00} = a_{11} = 1$, this means that SKE in the quantum game yields the payoff of 4, whereas playing SKE in the classical game results in the payoff of 2.5.
\end{enumerate}
\end{example}
\section{Conclusions}
Simple Kantian equilibria provide us with a prediction how a symmetric game might be played. It is based on the assumption that players choose the same strategies when they each face the same strategic position in the game. It greatly simplifies the analysis required to find a reasonable strategy profile as one comes down to finding a maximum of a function.  Our work has shown that the notion of SKE is suitable for both the classical and quantum games. By applying a positive affine transformation we simplified the structure of a symmetric $2\times 2$ game and derived a general formula of SKE in the classical game and possible payoff outcomes of SKE in the quantum game. We found that the result from playing SKE in the quantum game is at least as good as in the classical game. Moreover, in many cases, the players benefit from playing the quantum game getting a strictly higher payoff. 

Our work also aimed to show the advantages of non-Nashian solution concepts in quantum games. Pure Nash equilibria do not usually exist when the strategy sets are equal to $\mathsf{SU}(2)$ in the EWL model. If they can be found, the Nash equilibria are trivial in the sense that they determine payoffs equal to the Nash equilibrium payoffs in the classical game. In contrast, SKE always exist in the quantum approach to symmetric $2\times 2$ game which allows us to compare reasonable outcomes when the game is played in a classical and quantum manner. Our studies also initiate further research on symmetric games of higher dimension. Our future research will aim to examine $n$-person symmetric games as well as other non-Nashian solution concepts in quantum games.


\begin{thebibliography}{999}
\bibitem{neumann} von Neumann J., Zur Theorie der Gesellschaftsspiele, Mathematische Annalen 100 295 (1928).
\bibitem{neumann2} von Neumann J., Morgenstern O., Theory of Games and Economic Behavior, Princeton: Princeton University Press, (1944).
\bibitem{nash} Nash, J., Equilibrium points in $n$-person games, {\em Proc. Natl. Acad. Sci. U.S.A.} {\bf 36} 48 (1950).
\bibitem{myerson} Myerson R. B. Game Theory: Analysis of Conflict, Cambridge, Massachusetts: Harvard University Press.
\bibitem{fourny} Fourny G., Perfect Prediction in normal form: Superrational thinking extended to non-symmetric games, Journal of Mathematical Psychology 96, 102332 (2020)
\bibitem{berge} Berge C (1957) Th\'eorie g\'en\'erale des jeux \`a n personnes [General theory of $n$-person games], vol 138. Gauthier-Villars Paris
\bibitem{colman} Colman A.M, K\"orner T.W, Musy O, Tazda\"it  T (2011) Mutual support in games:
Some properties of Berge equilibria. Journal of Mathematical Psychology
55(2):166–175
\bibitem{pbf} Pykacz J., Bytner P., Fr\k{a}ckiewicz P. Example of a finite game with no Berge equilibria at all, Games 10(1), 7 (2019)
\bibitem{roemerbook} Roemer, J. E. (2019). How we cooperate: A theory of Kantian optimization. New Haven,
CT: Yale University Press.
\bibitem{hofstadter} Hofstadter D. (1983) Dilemmas for Superrational Thinkers, Leading Up to a
Luring Lottery. Scientific American
\bibitem{meyer} Meyer D. A., Quantum strategies, {\it Phys. Rev. Lett.} {\bf 82} 1052 (1999).
\bibitem{eisert} Eisert J., Wilkens M., Lewenstein M., Quantum games and quantum strategies, {\it Phys. Rev. Lett.} {\bf 83} 3077 (1999).
\bibitem{lidumassar} Li H., Du J., Massar S., Continuous-variable quantum games, {\it Phys Lett A} {\bf 306} 73 (2002).
\bibitem{fraduopol} Frąckiewicz, P. Quantum approach to Cournot-type competition. Int J Theor Phys 57, 353–362 (2018)
\bibitem{maszler} Maschler M., Solan E., and Zamir S., Game Theory, Cambridge University Press (2013).
\bibitem{binmore} Binmore, K. Playing For Real: A Text on Game Theory, Oxford University Press (2007)
\bibitem{osborne} Osborne M.J and Rubinstein A., A Course in Game Theory, MIT (1994)
\bibitem{plan} Plan, A., Symmetric $n$-player games, University of Arizona, Economics Working Paper 17-08.
\bibitem{roemer} Roemer, J. E. “A Theory of Cooperation in Games with an Application to Market Socialism” Review of Social Economy. doi: 10.1080/00346764.2018.1555647 (2019)
\bibitem{istrate} Istrate, G. Game-theoretic models of moral and other-regarding agents.
\end{thebibliography}
\end{document}